# Micro-structuration effects on local magneto-transport in [Co/Pd]IrMn thin films

C. Walker[1], M. Parkes[1], C. Olsson[1], D. Keavney[2], E. E. Fullerton[3] and K. Chesnel[1]

[1]Department of Physics and Astronomy, BYU, Provo, UT 84602, USA
[2]Advanced Photon Source, Argonne National Laboratory, Lemont, IL 60439, USA
[3]Center for Memory and Recording Research, U.C. San Diego, La Jolla, CA 92093-0401, USA

## ABSTRACT

We measured the local magneto-transport (MT) signal with an out-of-plane magnetic field, including magneto-resistance (MR) and Extraordinary Hall effect (EHE), in exchange-biased [Co/Pd]IrMn thin multilayers that are micro-structured with a 100 μm window. We found that when measured locally around the window, the MT signal deviate from the expected behavior. We studied possible causes, including film micro-structuration, electrical contact geometry as well as magnetic field tilt from the normal direction. These MT measurements were carried using the Van-der-Pauw method, with a set a four microscopic contacts directly surrounding the window, and a set of four contacts positioned several millimeters away from the window. We found that tilting the magnetic field direction with respect to the normal does not significantly affect the MT signal, whereas the positioning and geometry of the contacts seem to highly affect the MT signal. When the contacts are directly surrounding the window, the shape of the EHE signal is drastically deformed, suggesting that the electron path is disturbed by the presence of the window and the proximity of the electric contacts. If, on the other hand, the contacts are sufficiently far apart, the MT signal is not significantly affected by the presence of the window. Furthermore, the deformed EHE signal measured on the inner contacts can be modeled as a mix of the EHE and MR signals measured on the outer contacts.

## INTRODUCTION

Multilayered [Co/Pd]IrMn thin films, as sketched in Fig.1a, exhibit interesting magnetic properties.[1-3] One of these properties is exchange-bias (EB) [4], caused by interfacial couplings between the ferromagnetic (FM) Co/Pd multilayers and the antiferromagnetic (AF) IrMn layers, occurring when the film is field-cooled below its blocking temperature $T_B$. [5] We found in previous studies [6-8] carried on [[Co (4Å)/Pd (7Å)]$_{x12}$/IrMn(24Å)]$_{x4}$ for which $T_B \sim 275$ K, that these exchange couplings induce remarkable Magnetic Domain Memory (MDM). The observed MDM is the highest and the most extended throughout the magnetization loop when the cooling field is close to zero (remanence).[9] At remanence, the magnetic domains in the F layer tend to form a maze pattern, like the one illustrated in Fig. 1b, which gets imprinted in the AF layer upon cooling. We carried these studies using synchrotron x-ray radiation, which allows probing the domain pattern morphological changes at the nanoscale while applying a magnetic field in-situ.[10] However, recently, we observed an unexpected loss of MDM upon field cycling. We have been investigating possible reasons for the MDM loss, including x-ray illumination effects. For these investigations, we used in-situ magneto-transport (MT) while under x-ray illumination, to see if the MDM loss may be accompanied by a loss of EB. In practice, EB consists in the biasing of the magnetization loop with the respect to the applied magnetic field, in the direction opposite to the direction of the applied cooling field. Our earlier magnetization measurements on [[Co (4Å)/Pd (7Å)]$_{x12}$/IrMn(24Å)]$_{x4}$, shown in Fig.1c using Vibrating Sample Magnetometry (VSM), indicate a bias field as high as 200 Oe at 20 K. When measuring EB via MT, we could observe a biasing effect consistent with the VSM measurements. However, we found that the shape of the Extraordinary Hall Effect (EHE) signal (see Fig.1d) is deformed in respect to the expected magnetization loop shape. This paper investigates possible reasons for this deformation.

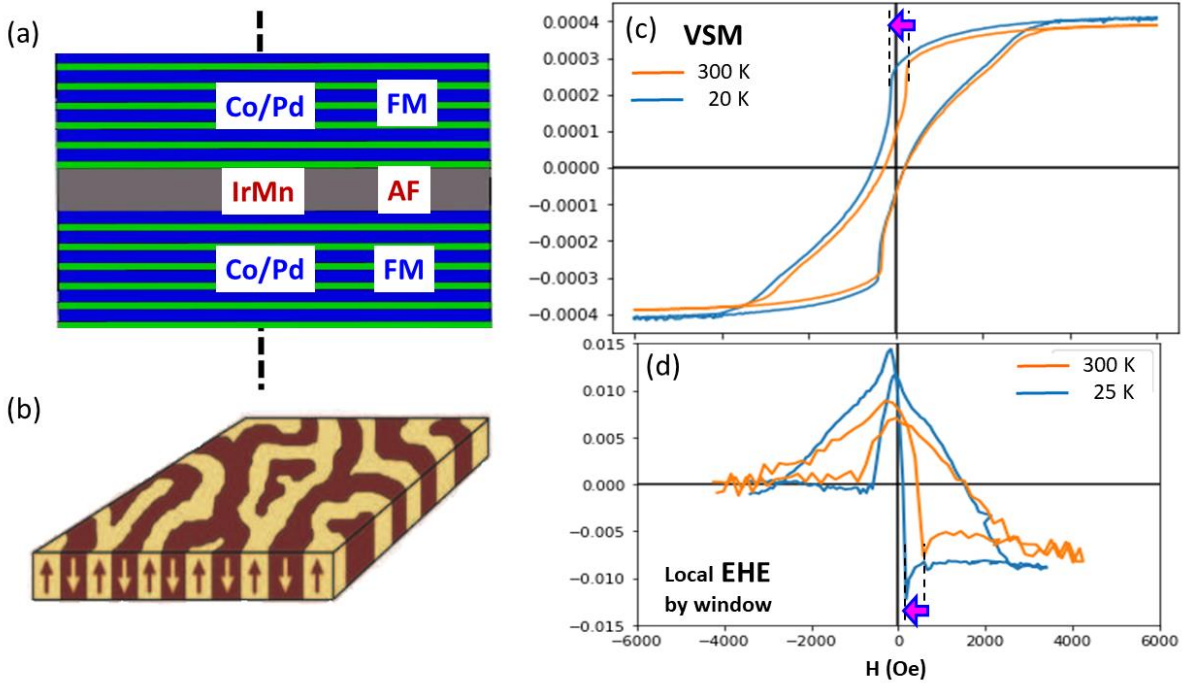

FIG.1 (a) Sketch of the [[Co (4Å)/Pd (7Å)]$_{x12}$/IrMn(24Å)]$_{x4}$ multilayer structure; (b) Illustration of magnetic domain pattern forming in the film near remanence; (c) Magnetization loops measured via VSM at 300 K and at 20 K after field cooling under a + 6000 Oe field; (d) Magnetization loops measured via EHE on inner contacts at 300 K and 25 K after field cooling under +4750 Oe, Both cooling fields were well above saturation point $H_s \sim$ 3200 Oe.

## METHODOLOGY

The synchrotron x-ray magnetic scattering measurements were carried at the Advanced Photon Source, beamline 4-ID-C, in a vacuum chamber equipped with an in-situ octupole magnet. To allow x-ray scattering measurements in transmission geometry, the [[Co (4Å)/Pd (7Å)]$_{x12}$/IrMn(24Å)]$_{x4}$ thin films were deposited onto 100 nm thick Si$_3$N$_4$ membranes supported by silicon wafers that have a 100 μm window at their center. To enable MT measurements, the films were electrically hooked to a circuit board with ultra-thin 20 μm wires welded via wire-bonding, as seen in Fig.2a. The electrical contacts were grouped by sets of four, to enable both Extraordinary Hall Effect (EHE) and magneto-resistance (MR) measurements using the Van-der-Pauw method [11, 12] with a magnetic field applied out-of-plane.

To study the effect of x-ray illumination on the exchange couplings and a possible loss of EB, the MT signal was measured locally, as close as possible to the illuminated window. For this purpose, we created four electrical contacts nearby the central window by depositing four Pt pads using Focused Ion Beam (FIB). The geometry and location of these four pads is shown in the sketch Fig.2b and in the Scanning Electron Microscopy (SEM) image Fig.2c. Each pad has a shape of a 125 x 125 μm$^2$ square with a 100 nm thickness. The four pads are diagonally located by the four corners of the window at a distance of 300 μm of each other. Along the diagonal, the distance between pads is about 425 μm center-to-center and the central window covers about 33% of that distance. Additionally to these micrometric "inner" contacts, we created four "outer" contacts located by the four corners of the wafer at about 5 mm of each other (see Fig.2b).

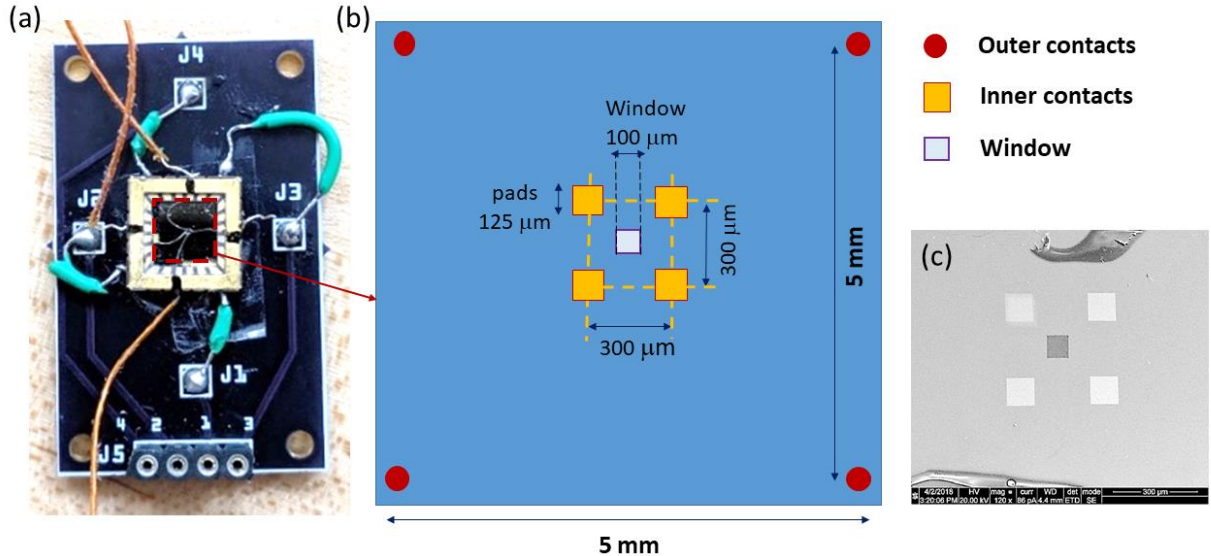

FIG.2 (a) Picture of the electrical board on which the film is mounted and electrically connected via wire-bonding; (b) Sketch showing the location of the outer contacts at the four corners of the film and the location of the inner contacts surrounding the central 100 μm window; (c) SEM image of the padded inner contact deposited via FIB surrounding the 100 μm window.

When measuring the MT signal on the inner contacts while using the octupole magnetic chamber, we found that the EHE signal, shown in Fig.1d, was deformed in respect to the expected hysteresis loop, as measured via VSM (see Fig.1c). The EHE signal appeared like a folded hysteresis loop.

To investigate the origin of this deformation, we conducted a series of MT measurements in our laboratory at BYU. One investigation consisted in comparing the MT signal on the inner contacts to the MT signal on the outer contacts, used as a reference. Another investigation consisted in studying possible effects caused by a tilt of the applied magnetic field with respect to the direction normal to the film surface. Indeed, during the synchrotron measurement, one of the eight poles of the octupole electromagnet failed, causing a tilting of the applied magnetic field with respect to the direction of the x-rays. In addition, due to space constraints, the sample holder was also tilted, causing a tilt of the film with respect to the x-ray direction. The combination of these two angular deviations resulted in a total tilt between the magnetic field direction and normal to the film surface up to about 25°.

The BYU MT measurements were carried in a bipolar electromagnet. The EHE signal was obtained by measuring the voltage in the transverse direction with respect to the applied current, while the magnetic field is applied perpendicular to the film. Averages between such transverse measurements measured at 90° of each other were plotted. The MR signal was obtained by measuring the voltage in the direction parallel to the applied current using the four contacts and applying the Van-der-Pauw method, where four different configurations are averaged to eliminate possible structural asymmetries. [12]

**RESULTS AND DISCUSSION**

The MT signal measured on the outer contacts, in Fig.3 a,b, shows a behavior typical of ferromagnetic materials. The averaged EHE signal in Fig.3a has the shape of a hysteresis loop, consistent with the VSM signal. The averaged MR signal in Fig.3b has a symmetrical double-lobe shape, typical of magneto-resistance in ferromagnetic thin films with perpendicular magnetic anisotropy.[13,14]

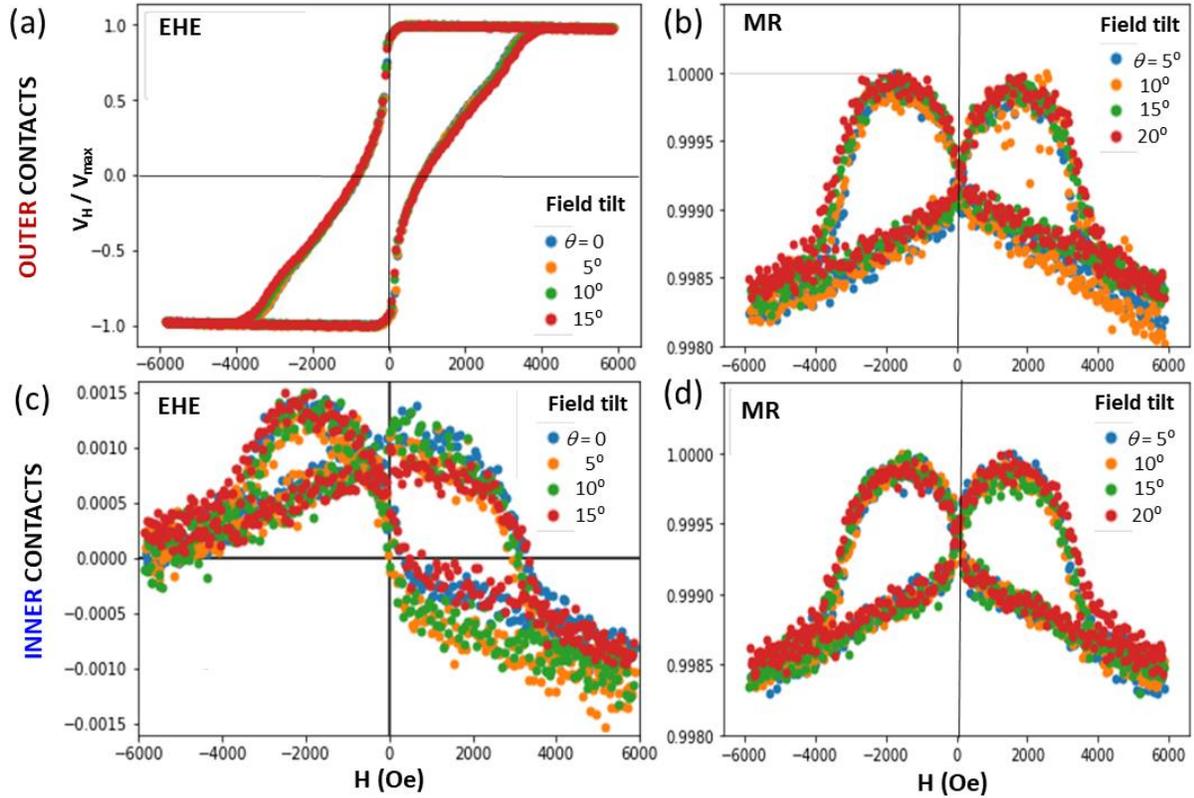

FIG.3 (a,b) MT signal measured on the *outer contacts:* (a) EHE signal and (b) MR signal; (c,d) MT signal measured on the *inner contacts*: (c) EHE signal and (d) MR signal, all measured at different angles from 0 to 20°.

The MT signal measured on the inner contacts however behaves differently compared to the MT signal measured on the outer contacts. The EHE signal in Fig.3c has a deformed shape with respect to the expected hysteresis loop. The deformed shape looks like a hysteresis loop folded onto itself in an asymmetrical way, similar to what was measured during the synchrotron experiment at the APS, in Fig.1d. The slight difference in shape between the APS and BYU measurement may be due to some of the micro-wires having disconnected and new micro-wires being welded at distinct locations. On the other hand, the average MR signal in Fig.3d is similar to the MR signal measured on the outer contacts (Fig.3b), still showing a symmetrical double-lobe shape.

The dependence with magnetic field tilt was studied by tilting the sample holder with respect to the electromagnet axis. The setup allowed a tilt up to 20°. The data plotted in Fig.3 shows no significant effect of magnetic field tilt on the shape of the EHE and MR signals, neither on the outer contacts nor on the inner contacts. This observation rules out any correlation between the observed EHE signal deformation and the actual magnetic field tilt during the synchrotron measurement.

It then appears that the deformation of the EHE signal observed on the inner contacts is principally due to micro-structuration effects and geometry of the contacts. When measuring the MT signal using the inner contacts, the electron path is significantly disturbed by the presence of the window, which occupies 33% of the distance between contacts, and also by the relative width of the pads (125 μm) with respect to the inter-pad distance (300 μm), that is about 45 %. This suggests that electrons may not travel on straight paths between diagonally opposite contacts, but instead may deviate from the straight path to get around the central window. In this process, some electrons may end up hitting adjacent contacts instead of opposite contacts, causing some mixing between EHE and MR signals.

To support this hypothesis, we attempted modeling the EHE signal measured on inner contacts by using a linear combination of the EHE and MR signals measured on the outer contacts as follows:

$$(EHE)_{IN} = a * (EHE)_{OUT} + b * (MR)_{OUT}$$

The modeling results in Fig. 4 show that one can indeed reconstruct the asymmetric shape of $(EHE)_{IN}$ by mixing the $(EHE)_{OUT}$ and $(MR)_{OUT}$ signals with coefficients $a$ and $b$ in opposite signs. The ratio $|b/a|$ ranges between 1.5 and 1.7 for the various sets measured at different tilt angles.

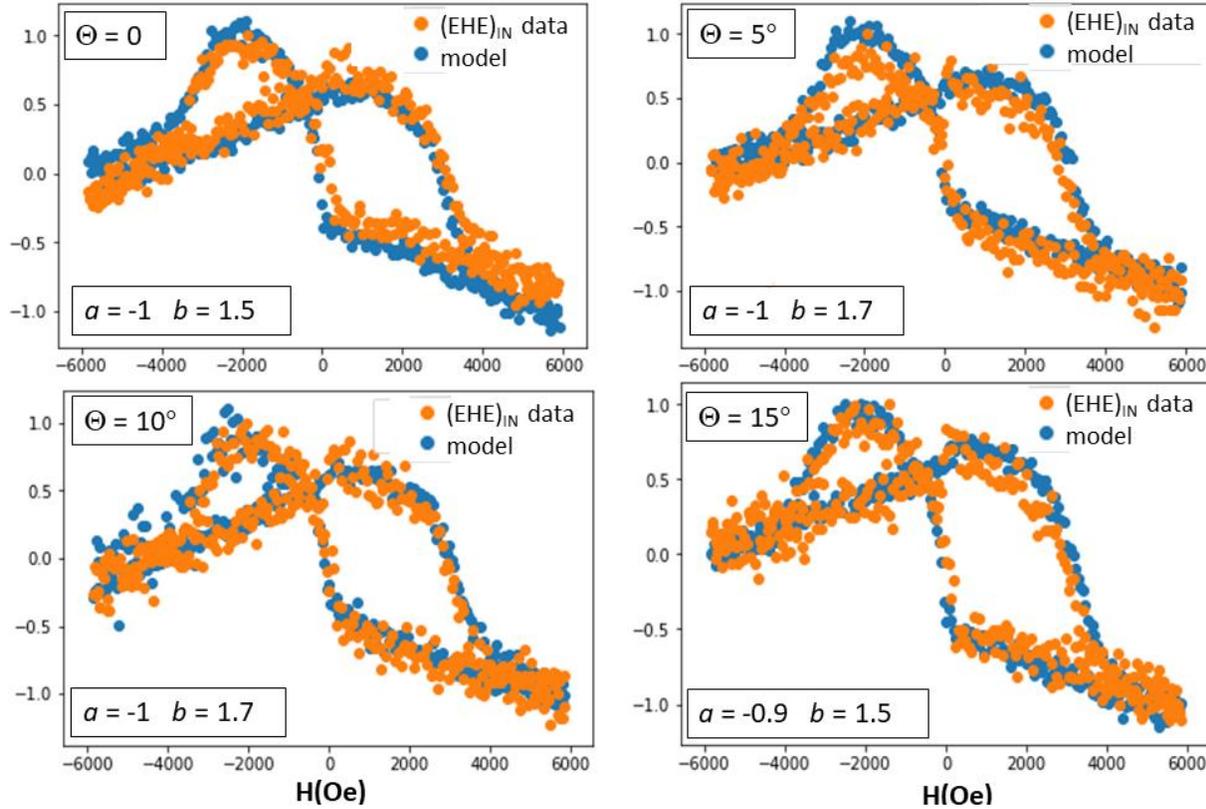

FIG.4 Modeling of the EHE signal measured on the inner contacts by using a linear combination of the EHE and MR reference signals measured on the outer contacts, as follows: $(EHE)_{IN} = a * (EHE)_{OUT} + b * (MR)_{OUT}$

## CONCLUSION

We have probed local magneto-transport (MT) signal, including magneto-resistance (MR) and Extraordinary Hall Effect (EHE), in exchange-biased [[Co (4Å)/Pd (7Å)]$_{x12}$/IrMn(24Å)]$_{x4}$ thin films that are micro-structured with a central 100 μm window. We carried these measurements using the Van-der-Pauw method on two sets of four contacts: a set of outer contacts located 5 mm apart at the four corners of the film; and a set of inner contacts, made of 125 μm pads located at 300 μm of each other, surrounding the central 100 μm window. We found that when measured on the outer contacts, the MT signal has the expected shape, with the EHE signal forming a hysteresis loop consistent with magnetometry measurements, and the MR signal showing a symmetrical double-lobe shape. When measured on the inner contacts, the MR still shows the symmetrical double-lobe shape, however the EHE signal is significantly deformed, looking like an asymmetric folded hysteresis loop. This deformed shape may be reconstructed by mixing the EHE and MR signals measured on the outer contacts, using a relative weight ratio for MR/EHE in the range of 1.5 to 1.7, with opposite signs. This suggests that when EHE is probed locally around the window, the electrons are not traveling in a straight path between diagonally opposite contacts but a portion of them hit the adjacent contacts instead, leading to a mix of MR and EHE signals. Additionally, we found that a tilt of the magnetic field with respect to the film normal direction is not affecting the shape of the MT signal significantly. The observed deformation of the EHE signal is mainly due to the micro-structuration of the film and the proximity of the inner contacts with the central window covering 33% of the path between contacts.

In order to measure effects of x-ray illumination on exchange-bias, it is however necessary to probe the MT signal locally around the illuminated region of the film, which necessitates some micro-structuration. Our current setup induces a deformation of the EHE signal. A solution to this issue may be to etch the [Co/Pd]IrMn film in the shape of a cross around the central window so to guide the electrons path in the desired direction for the EHE measurement. That being said, it is interesting to note that the EHE signal measured on the inner contacts with the current setup still shows biasing effect when the film is cooled from 300 K down to 20 K below the blocking temperature, consistent with magnetometry measurements. So, even when its shape is deformed due to micro-structuration, the EHE signal may be used to measure exchange bias and monitor its dependence with various parameters such as temperature or x-ray illumination.

**Acknowledgment**

This work was supported by the Research Experience for Undergraduate Students (REU) funding program, grant #1757998, at the National Science Foundation (NSF) as well as by the Advanced Photon Source at Argonne National Laboratory, a Department of Energy facility.

**Data Availability**

The data that support the findings of this study are available from the corresponding author upon reasonable request.